\documentclass[aps,preprint,showpacs,amsmath,amssymb,prd,nofootinbib]{revtex4}
\usepackage{amssymb,graphics,graphicx}
\newcommand{\be}{\begin{equation}} \newcommand{\ee}{\end{equation}}
\newcommand{\bea}{\begin{eqnarray}}
\newcommand{\eea}{\end{eqnarray}}

\begin{document}
\title
{\large \bf Three Generations in Minimally Extended Standard Models}
\author{Paul H. Frampton$^{1,2}$\footnote{paul.h.frampton@gmail.com}, Chiu Man
Ho$^{3}$\footnote{chiuman.ho@vanderbilt.edu}, and Thomas
W. Kephart$^{3}$\footnote{tom.kephart@gmail.com\\ $^\#$ permanent address}}
\affiliation{$^{1}$Department of Physics and Astronomy, University of North Carolina,
Chapel Hill$^\#$, NC 27599, USA\\
$^{2}$Centro Universitario DeVoto, Villa DeVoto, Buenos Aires, Argentina  \\
$^{3}$Department of Physics and Astronomy, Vanderbilt University, Nashville, TN 37235, USA}
\date{\today}

\begin{abstract}
We present a class of
minimally extended standard models
with the gauge group $SU(3)_C \times SU(N)_L \times U(1)_X$ where for all $N \geq 3$,
anomaly cancelation requires three generations.
At low energy, we recover the
Standard Model (SM), while at higher
energies, there must exist quarks, leptons and gauge bosons
with electric charges shifted from their SM values by integer multiples of the electron charge up to $ \pm\, [N/2]\,e$.
Since the value $N=5$ is the highest $N$ consistent with QCD asymptotic freedom,  we elaborate on the 3--5--1 model.
\end{abstract}

\maketitle

\section{Introduction}

With expected availability of data in $pp$-scattering
up to 14 TeV center-of-mass energy in the next
few years at the Large Hadron Collider
(LHC), it is reasonable to anticipate additional gauge bosons as well as
more chiral fermions beyond the Standard Model (SM).
We expect the Standard Model to be the low energy limit of any new physics seen at the LHC.
Hence, whatever the new physics, it should be an extension of the Standard Model. In order to
explore the parameter space of such models and at the same time make testable predictions for the LHC,
it is best to consider minimal extensions of the Standard Model, since these
have the best hope of being testable and they are likely to be contained in the ultimate model.

Merely adding sequential quarks and leptons beyond the first
three generations, while keeping the Standard Model $SU(3)_C \times SU(2)_L \times U(1)_Y$ gauge group,
is the simplest possibility, but perhaps
not the most attractive scenario based on purely aesthetic grounds.
For instance, eight families are the most allowed by QCD asymptotic freedom.
But the symmetry requirement for baryons \cite{Greenberg:1964pe} seems to suggest that the number of
colors should remain three. Thus, a somewhat more elegant model would have the overall gauge group grow
in rank at higher energy while the color gauge group remains fixed.  In this approach, only the chiral electroweak
gauge group is extended from rank 2 to rank $N$ for $N \geq 3$, namely we consider $SU(2)_L \longrightarrow
SU(N)_L$. There will remain a $U(1)_X$ factor in the non-simple gauge group. For
the Standard Model, we have $X = Y$ (the weak hypercharge) which cancels all triangle
anomalies, generation by generation.
For $N \geq 3$, however, $X$, though still unpredicted, becomes notably
simpler than $Y$.

Following the above strategy, it is shown here how to generalize the SM gauge group to a
minimally extended standard model (MESM)
with gauge group $SU(3)_C \times SU(N)_L \times U(1)_X$ such that for all
integers $N \geq 3$, unlike the SM case, anomaly cancelation requires three generations.
At low energy, the models reduce to the
Standard Model. At higher
energies, there must exist quarks and leptons
with electric charges up to $Q = q \pm \,[N/2]\,e$ where $q$ is the corresponding SM charge and the symbol $[x]$ denotes the integer
part of $x$, together with gauge bosons which have electric charges
up to $Q=  \pm\, [N/2 + 1]\,e$.

\section{Minimally Extended Standard Models}

Our construction is inspired
at the first stage by the 331-model \cite{ Frampton:1992wt,Pisano:1991ee} but involves generalization
to give an infinite sequence of three-generation anomaly-free extensions. (For reviews and related work, see \cite{Ng:1992st,Cuypers:1996ia,Dimopoulos:2002mv,Pleitez:1994pu}.)

It is easier to start with odd $N = (2n +1)$, then describe even $N = 2n$.
For $N = (2n + 1)$, the first and second generations of fermions are assigned under $(\, SU(3)_C, \;SU(N)_L\,)_X$ as
\begin{equation}
(\,3, \,N\,)_{-\frac{1}{3}}\, +\, (\,\bar{3},\, 1\,)_{+\frac{1}{3}}\,+ \,(\,\bar{3},\, 1\,)_{+\frac{1}{3}\pm 1}\,+ ...+\,(\,\bar{3},\, 1\,)_{+\frac{1}{3} \pm n}\,
+\, (\,1, \,\bar{N}\,)_0 \,+\, (\,1,\, 1\,)_0\,.
\label{first}
\end{equation}
More explicitly, the first two generations of quarks are assigned in \eqref{first} as the $\mathbf{N}$ of $SU(N)_L$:
\begin{equation}
Q_{iL}= \left( \begin{array}{c}
U_i^{n-1} \\
U_i^{n-2} \\
.\\
.\\
U_i^{1}\\
u_i \\
d_i\\
D_i^{-1}\\
.\\
.\\
D_i^{-(n-1)}\\
D_i^{-n}
\end{array}
\right)_L, ~~ i=1,2 ~; ~~~~~~ X=-\frac{1}{3}\,,
\label{1stN}
\end{equation}
where the notations $U_i^{n-1}$ and $D^{-n}$ mean that the charge is shifted up by $(n-1)\,e$ units and down by $-n\,e$ units respectively.
The first two generations of leptons are assigned in \eqref{first} as the $\mathbf{\bar{N}}$ of $SU(N)_L$:
\begin{equation}
f_{lL}=\left( \begin{array}{c}
E_l^{-n} \\
E_l^{-(n-1)} \\
.\\
.\\
E_l^{-2}\\
l^{-} \\
\nu_l\\
l^{+}\\
E_l^{2}\\
.\\
. \\
E_l^{n-1}\\
E_l^{n}
\end{array}
\right)_L, ~~ l=e,\mu ~; ~~~~~~ X=0\,.
\label{1stbarN}
\end{equation}

We shall write triangle anomalies in the order
\begin{equation}
A_g = [\, SU(3)_C^3,\; SU(3)_C^2\,X,\; SU(N)_L^3,\; SU(N)_L^2\,X,\; X^3,\; X\,]\,,
\label{triangle}
\end{equation}
where $g=1,2,3$ \,is the generation number. The sixth and last anomaly is
a mixed gravity-gauge triangle diagram.
For $N\geq 3$, normalizing the defining representation of $SU(N)$ to have anomaly $+1$
and normalizing $X$ as defined above, we find that the first two generations of fermions
have anomalies
\begin{equation}
A_1 = A_2 = [ \,0,\; 0,\; +2,\; -1,\; n\,(n+1)\,(2n+1),\; 0\,]\,.
\label{anoms1}
\end{equation}

The third generation of fermions is assigned under $(\, SU(3)_C, \;SU(N)_L\,)_X$ as
\begin{equation}
(\,3, \,\bar{N}\,)_{+\frac{2}{3}} \,+\, (\,\bar{3},\, 1\,)_{-\frac{2}{3}}\,+\,(\,\bar{3},\, 1\,)_{-\frac{2}{3} \pm 1}\,+...+\,(\,\bar{3},\, 1\,)_{-\frac{2}{3} \pm n} \,+\,
(\,1, \,\bar{N}\,)_0\, + \,(\,1,\, 1)_0 \,.
\label{third}
\end{equation}

More explicitly, the third generation of quarks is assigned in \eqref{third} as the $\mathbf{\bar{N}}$ of $SU(N)_L$:
\begin{equation}
Q_{3L}=\left( \begin{array}{c}
B^{-(n-1)} \\
B^{-(n-2)} \\
.\\
.\\
B^{-1}\\
b \\
t\\
T^{1}\\
.\\
.\\
T^{n-1}\\
T^{n}
\end{array}
\right)_L, ~~~~~~ X =\frac{2}{3}\,.
\label{3rdbarN}
\end{equation}
Similar to the first two generations of leptons, the third generation of leptons is also assigned in \eqref{third} as the $\mathbf{\bar{N}}$ of $SU(N)_L$. Thus, the index $l$ in \eqref{1stbarN} can actually be\, $e, \mu$ or $\tau$. The anomalies due to the third generation of fermions are
given by
\begin{equation}
A_3 = [\, 0, \;0, \;-4, \;+2, \;-\,2\,n\,(n+1)\,(2n+1), \;0\,]\,.
\label{anoms3}
\end{equation}

Combining \eqref{anoms1} and \eqref{anoms3} leads to
\begin{equation}
A_1 + A_2 + A_3 = [ \,0, \;0, \;0, \;0, \;0, \;0\,]\,.
\label{anoms123}
\end{equation}
Thus, the anomalies from the first two generations of fermions exactly cancel those from the third generation of
fermions. There are many ways to express the cancelation physically, perhaps the
simplest is to say that the number of generations must equal to the number of
colors.

For even $N=2n$ \cite{Foot:1994ym}, the simplest possibility is to imitate the Standard Model with sequential generations where the required extension is most easily seen by the breaking of $SU(2n+1)$ to $SU(2n)$ which cannot generate new anomalies. In the $SU(2n)$ models, the leptons develop non-zero $X$ charges but, as expected, the six anomalies listed in \eqref{triangle}
all cancel for three generations. It is straightforward to check, that unlike the minimal odd $N$ models, anomalies cancel generation by generation for even $N$ as in the Standard Model.
In breaking $SU(2n+1)$ to $SU(2n)$, there are two possibilities
for $SU(2n)$. The reason is that we can assign the singlet to the fundamental irrepresentation in the natural embedding
$\mathbf{N} = (\mathbf{N}-\mathbf{1})+  \mathbf{1}$ to either the highest weight
or the lowest weight of the $\mathbf{N}$.

As a final remark, for a given $N$, the models with three generations have 3\,$N$ quark flavors, and so we require $N \leq 5$
for QCD asymptotic freedom, so we will turn our focus to this 3--5--1 model. While the $N = 5$ model is close to losing asymptotic freedom, it is also close to being conformal at one loop, an interesting feature that we will not explore here \cite{Hung:2009ia,Hung:2009hy,Ho:2011qi}.

\section{Higgs Sector and Symmetry Breaking}

We expect the effective nonabelian part of the electroweak gauge group to be N = 3, 4, ... as the energy
increases. As shown in \cite{Frampton:1992wt}, the $SU(3)_L$ breaking must be below
$\sim 4$ TeV for reasons associated with the running of $SU(2)_L$ coupling constant
and the embedding of $SU(2)_L$ in $SU(3)_L$. The corresponding
new gauge bosons $(Y^{++}, Y^+), (Y^{--},Y^-)$ and $Z^{'}$, as well
as the exotic quarks $D\, (\, Q=-4/3\,)$ and $T\,(\, Q=+5/3\,)$ are
predicted to have masses $\lesssim 4$ TeV. For higher $N\geq 4$,
the scale of symmetry breaking for $SU(N)_L \rightarrow SU(N-1)_L$
will be successively higher.

As an illustrative example, we consider a model with the gauge group $SU(3)_C \times SU(5)_L \times U(1)_X$. As mentioned earlier, this
3-5-1 model is the maximal model that still preserves the QCD asymptotic freedom. It has the following fermion contents:
\bea
f_{lL} &=& \left(
  \begin{array}{c}
    E_l^{--} \\
    l^{-} \\
     \nu_l\\
    l^{+} \\
    E_l^{++} \\
  \end{array}
\right), ~~~~ l = e, \mu, \tau \,, \\
Q_{iL} &=&\left(
  \begin{array}{c}
    U_i^{1} \\
    u_i \\
    d_i\\
    D_i^{-1} \\
    D_i^{-2} \\
  \end{array}
\right), ~~ i=1,2 ~~~~~~~~
Q_{3L}=\left(
  \begin{array}{c}
    B^{-1} \\
    b \\
    t\\
    T^{1} \\
    T^{2} \\
  \end{array}
\right)\,,
\eea
where we have denoted $E_l^{-2}$ and $E_l^{2}$\, by $E_l^{--}$ and $E_l^{++}$ respectively.

The symmetry breaking from $SU(3)_C \times SU(5)_L \times U(1)_X$ to $SU(3)_C \times SU(2)_L \times U(1)_Y$ (where $Y$ is the hypercharge)
can be achieved with three fundamental Higgs fields $\phi,\,\chi$ and $\Delta$ acquiring the following vacuum expectation values (VEVs):
\bea
\langle \phi \rangle = \left(
  \begin{array}{c}
    0 \\
    0 \\
    0\\
    0 \\
    v \\
  \end{array}
\right),~~~
\langle \chi \rangle = \left(
  \begin{array}{c}
    v' \\
    0 \\
    0\\
    0 \\
    0 \\
  \end{array}
\right),~~~
\langle \Delta \rangle = \left(
  \begin{array}{c}
    0 \\
    0 \\
    0\\
    v'' \\
    0 \\
  \end{array}
\right)\,.~~~
\eea
These VEVs can be obtained by constructing the following Higgs potential:
\bea
V(\phi,\chi,\Delta) &=& \alpha_1\,\left(\phi^\dagger \phi - v^2\right)^2 + \alpha_2\,\left(\chi^\dagger \chi - v'^{\,2}\right)^2
+\alpha_3\,\left(\Delta^\dagger \Delta - v''^{\,2}\right)^2 \nonumber \\
&& + \alpha_4\,\left[ \left(\phi^\dagger \phi  - v^{\,2}\right) + \left(\chi^\dagger \chi - v'^{\,2}\right) \right]^2
+ \alpha_5\,\left[ \left(\phi^\dagger \phi  - v^{\,2}\right) + \left(\Delta^\dagger \Delta - v''^{\,2}\right) \right]^2
\nonumber \\
&& + \alpha_6\,\left[ \left(\chi^\dagger \chi  - v'^{\,2}\right) + \left(\Delta^\dagger \Delta - v''^{\,2}\right) \right]^2\,,
\eea
where all the coefficients $\alpha_1, ...,\alpha_6$ are non-negative. Obviously, these VEVs minimize the Higgs potential.

After the symmetry breaking, we can  find the expression for the hypercharge $Y$. With the convention
$ \rm{Tr} (\lambda_a\,\lambda_b) = 2\, \delta_{ab}$, we adopt the following representation for the generators (note our $\lambda$ matrices differ from the standard notation by a reordering of rows and columns):
\bea
T_3 &=& \frac12\,\rm{diag}\,\left(0, 1, -1, 0, 0\right)\,, \\
\lambda_8 &=& \frac{1}{\sqrt{3}}\, \rm{diag}\,\left(0, 1, 1, -2, 0\right)\,, \\
\lambda_{15} &=& \frac{1}{\sqrt{6}}\, \rm{diag}\,\left(0, 1, 1, 1, -3\right)\,, \\
\lambda_{24} &=& \frac{1}{\sqrt{10}}\, \rm{diag}\,\left(-4, 1, 1, 1, 1\right)\,.
\eea
It is then straightforward to show that the hypercharge $Y$ is given by
\bea
\label{Y}
Y= 2\,X + \sqrt{3}\;\lambda_{8}+ \sqrt{6}\;\lambda_{15} -\sqrt{10}\;\lambda_{24}\,,
\eea
with $X=0, \,X=-1/3$ and $X=2/3$\, for the leptons and quarks respectively. This expression for hypercharge $Y$ holds given that the
Higgs fields $\phi,\,\chi$ and $\Delta$ \,carry the $U(1)$ charges $X=2, \,X=-2$ and $X=1$\, respectively.
One can readily verify that \eqref{Y}
precisely reproduces the SM hypercharges for those leptons and quarks.
The electric charge $Q$ is given by
\bea
Q= T_{3} + \frac{Y}{2}\,,
\eea
which again reproduces the SM electric charges for the leptons and quarks of the standard model sector. Moreover, this expression for $Q$
also generates the required electric charges for $E^{--}, E^{++}, U_i^1, D_i^{-1,-2}, B^{-1}$ and $T^{1,2}$.

\section{Discussions}

Preliminary data from the LHC give no encouragement to
the popular ideas of supersymmetry and extra dimensions. So it is timely
to ask what are other extensions of the Standard Model which stay in four
bosonic space dimensions.

We have presented minimally extended standard models (MESMs).
They can be designated as {\it minimal} because the extension of
$SU(2)_L$ to $SU(N)_L$ with some simplification
of the $U(1)_X$ assignments compared to $U(1)_Y$, as we have noted, is surely
the infinite sequence which most simply extends the Standard Model.

The phenomenological consequences of the MESMs are spectacular.
It predicts that fermions and gauge bosons of ever increasing
electric charge will appear as the available energy increases.
It is unnecessary for $N \rightarrow \infty$ because of the
intervention of gravity.
Also, for electric charges near $\alpha^{-1}\,e$, the Schwinger mechanism \cite{Schwinger:1951nm} destabilizes
the vacuum, which provides a finite bound on $N$. (The interplay between the Schwinger effect and the implied large $N$
expansion   \cite{'tHooft:1973jz,Witten:1980sp,Di Vecchia:1980ve} of the electroweak sector could be an interesting avenue to explore.) However, there is no theoretical reason not to have gauge bosons,
hadrons, leptons etc., with charges $\pm2\,e, \,\pm3\,e, ....$; or quarks with large fractional charges.
At the LHC, we expect to see at least
the quarks with charges $Q=-4/3, \,+5/3$ and the doubly-charged
bileptons $Y^{\pm\,\pm}$, both corresponding to the new level $N=3$.
It is possible that the higher levels $N\geq4$
will be in evidence too. For example, in the case $N=5$, the highest $N$ allowed by QCD asymptotic freedom, there
are gauge bosons with charges up to $\pm4\,e$ coupling to channels like $\mu^- \mu^- \mu^- \mu^-$ which can provide
striking signatures at the LHC \cite{FHK2}.

\begin{acknowledgments}
The work of P.H.F. was supported in part by U.S.
Department of Energy Grant No. DE-FG02-05ER41418.
C.M.H. and T.W.K. were supported by U.S. DoE Grant
No. DE-FG05-85ER40226.
\end{acknowledgments}

\end{document}